\documentclass[%
aps,
prb, 
reprint, 
unsortedaddress,
superscriptaddress, 
letterpaper, 
showpacs, 
showkeys, 
amsmath,amssymb,amsfonts,
nobibnotes, 
floatfix, 
]{revtex4-1}

\usepackage[dvips]{graphicx} 
\usepackage[latin1]{inputenc}
\usepackage{dcolumn}
\usepackage{bm}
\usepackage{empheq} 

\usepackage{color}

\makeatletter
\let\oldtheequation\theequation
\renewcommand\tagform@[1]{\maketag@@@{\ignorespaces#1\unskip\@@italiccorr}}
\renewcommand\theequation{(\oldtheequation)}
\makeatother

\usepackage[colorlinks=true,
allcolors=blue,
linktocpage=true,
bookmarks=true,
bookmarksnumbered=true,
hyperindex,
breaklinks=true, 
pdfstartpage=1,%
pdfcenterwindow=false,%
pdffitwindow=true,%
pdfstartview={FitV},%
bookmarks=true,%
pdftitle={Temperature of a nanoparticle above a substrate under radiative heating and cooling},
pdfauthor={Houssem Kallel, Remi carminati, and Karl Joulain},
pdfcreator={LaTeX with hyperref package},
pdfproducer={dvips + ps2pdf},
pdfdisplaydoctitle=true,
dvips, 
ps2pdf,
]{hyperref} 
\usepackage[hyphenbreaks]{breakurl} 

\setcitestyle{numbers,square,comma,sort}

\begin{document}


\title{Temperature of a nanoparticle above a substrate under radiative heating and cooling}
\author{\firstname{Houssem} \surname{Kallel}}
\email{houssem.kallel@univ-poitiers.fr}
\email{houssem.kallel@yahoo.fr}
\affiliation{Institut Pprime, CNRS, Universit\'{e} de Poitiers, ISAE-ENSMA, F-86962 Futuroscope Chasseneuil, France}
\author{\firstname{R\'{e}mi} \surname{Carminati}}
\email{remi.carminati@espci.fr}
\affiliation{Institut Langevin, ESPCI Paris, PSL Research University, CNRS, 1 rue Jussieu, F-75005, Paris, France}
\author{\firstname{Karl} \surname{Joulain}}
\email{karl.joulain@univ-poitiers.fr}
\affiliation{Institut Pprime, CNRS, Universit\'{e} de Poitiers, ISAE-ENSMA, F-86962 Futuroscope Chasseneuil, France} 

\date{\today}

\begin{abstract}

Controlling the temperature in architectures involving nanoparticles and substrates is a key issue for applications involving micro and nanoscale heat transfer. We study the thermal behavior of a single nanoparticle interacting with a flat substrate under external monochromatic illumination, and with thermal radiation as the unique heat loss channel. We develop a model to compute the temperature of the nanoparticle, based on an effective dipole-polarizability approach. Using numerical simulations, we thoroughly investigate the impacts of various parameters affecting the nanoparticle temperature, such as the nanoparticle-to-substrate gap distance, the incident light wavelength and polarization, or the material resonances. This study provides a tool for the thermal characterization and design of micro or nanoscale systems coupling substrates with nanoparticles or optical antennas. 

\end{abstract}





\maketitle

\section{Introduction}

The impact of a Continuous-Wave (CW) illumination on the temperature of a nanoparticle (NP) has been the subject of much research in the last decade~\cite{Govorov2006,Baffou2011,Carlson2011,Setoura2012}. These researches consolidate the actual efforts aiming to evaluate and/or to optimize the potential use of NPs for photothermal therapy or photothermal imaging~[\onlinecite{Boisselier2009},\onlinecite{Jaque2014}], and light-assisted nanomaterial growth~[\onlinecite{Cao2007}].

The illumination obviously causes heat deposition inside the NP. Thermal radiation is a heat loss channel to be considered in addition to thermal conduction and thermal convection. This heat loss channel is influenced by spectral resonances, as experimentally verified, similar to those present in absorption or scattering ~spectra~[\onlinecite{Schuller2009}]. A super-Planckian emission behavior is also theoretically speculated to occur for an individual NP at the resonance wavelength~[\onlinecite{Centini2015}]. 

The radiative heat transfer between a single nano/micro-particle and a plane surface has been extensively investigated~[\onlinecite{Mulet2001,Narayanaswamy2008,Chapuis20081,Biehs2010,Tschikin2012,Miller2015}]. The scattering of light by a nano/micro-particle placed in front of a substrate has also attracted particular interest~[\onlinecite{Joulain2014}]. This has helped in understanding the signals obtained with thermal infrared near-field spectroscopy (TINS)~[\onlinecite{Jones2012}] and thermal radiation scanning tunneling microscopy (TRSTM)~[\onlinecite{Babuty2013}]. In these experimental techniques, a tip interacting with the surface scatters the evanescent electromagnetic waves.

The heat generated inside the NP by an external illumination scales with the ability of the NP to absorb the incoming light. The knowledge of the absorption efficiency of the NP is necessary for such kind of problem. Diverse numerical electromagnetic methods are suitable for the computation of the optical response of a single NP in the presence of a substrate such as the Discrete Dipole Approximation with Surface Interaction (DDA-SI)~[\onlinecite{Loke2011,Yurkin2015,Edalatpour2016}], the Boundary Element Method (BEM)~[\onlinecite{Waxenegger2015}], and the generalized Mie theory (T-matrix method or exact multipole expansion method)~
\cite{Videen1991,*Videen1992,*Videen1993,*Videen1995,*Videen2000,Fucile1997,Wriedt1998,Mackowski2008,Lerme2013}. 
Approximate methods, with considerably less computational resource requirements, can also be used, for example, the image dipole approach~[\onlinecite{Hillenbrand2002}] and the effective (or dressed) polarizability approach~\cite{Joulain2014,Miroshnichenko2015}.

The main objective of this work is to study quantitatively the influence of a CW-illumination on the temperature of a NP above a substrate in the presence of radiative heat losses from the NP to its surrounding medium, which, ideally, should be the vacuum. We develop an effective polarizability approach to evaluate the absorption efficiency of a NP interacting with a flat surface. We analyze different parameters that can influence the temperature of the NP, and discuss two kinds of NPs which substantially differ in terms of their optical absorption.

The paper is organized as follows. We describe the problem and detail the procedure used for the calculation of the NP temperature in Sec.~\ref{sec:Method}. In Sec.~\ref{sec:Results} we analyze the absorption efficiency of the NP, present numerical calculations of its temperature, and investigate the influence of the substrate and the NP material. In Sec.~\ref{sec:Results} we compute the intensity of the external illumination that produces a heating of the NP up to its melting point. Finally, we summarize the main results in Sec.~\ref{sec:concl}.  

\section{Statement of the problem and solution method}
\label{sec:Method}

\begin{figure}[!h]
\centering
\includegraphics*[width=8cm]{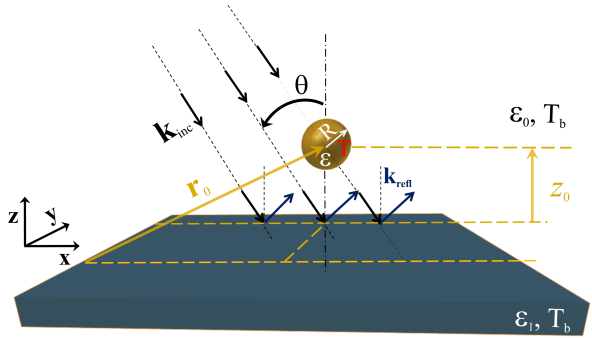}
\caption{Geometry of the system and definition of the parameters. The NP placed in front of a flat substrate is heated by an external plane wave, and exchanges heat with the environment at temperature \textit{T}$_{\textrm{\tiny{b}}}$ through radiation channel only.}
\label{fig1}
\end{figure}

We study a system made of an individual spherical NP of radius \textit{R}, located at a distance \textit{z}$_{\textrm{\tiny{0}}}$ from the surface of a flat substrate, as shown in Fig.~\ref{fig1}. Heat transfer between the NP and its environment is assumed to be dominated by radiation, which occurs for a NP in a vacuum (or in a gas at sufficiently low pressure~\cite{Narayanaswamy2008}), the environment being connected to a thermal bath of temperature \textit{T}$_{\textrm{\tiny{b}}}$. The NP is illuminated in steady-state by an external laser beam, modelled as a monochromatic plane wave with wavelength $\lambda$. The incident wave vector \textbf{k}$_{\mathrm{inc}}$ lies in the \textit{xz} plane, and makes an angle $\theta$ with respect to the normal of the substrate surface.

Our goal is to compute the temperature \textit{T} of the NP in a steady-state. We assume that the temperature inside the NP is uniform. Refering to the principle of energy conservation, the power absorbed by the NP equals the difference between the power $P_{\mathrm{rad}}\left(T\right)$ thermally emitted  by the NP at temperature  \textit{T} and the power emitted by the NP at the bath temperature \textit{T}$_{\textrm{\tiny{b}}}$:
\begin{equation}
P_{\mathrm{abs}}=~P_{\mathrm{rad}}\left(T\right)~-~P_{\mathrm{rad}}\left(T_{\textrm{{\tiny{b}}}}\right).
\label{eq:energy_conservation}
\end{equation}
The power absorbed by the NP is connected to the incident power carried by the external beam by the relation
\begin{equation}
P_{\mathrm{abs}}=~Q_{\mathrm{abs}}\left(\lambda,T\right)~I_{\mathrm{inc}}~C_{{\textrm{\tiny{geo}}}}
\label{eq:Pabs}
\end{equation}
where \textit{Q}$_{\mathrm{abs}}$ is the absorption efficiency, a dimensionless quantity defined as the ratio of the absorption cross-section \textit{C}$_{\textrm{abs}}$  and the geometrical cross-section \textit{C}$_{\textrm{\tiny{geo}}}=~\pi$\textit{R}$^{\textrm{{\tiny$2$}}}$ of the NP, and \textit{I}$_{\mathrm{inc}}$ is the intensity (power density or irradiance) of the external light beam. The power thermally emitted by the NP at temperature \textit{T} is~[\onlinecite{Bohren2008}]:
\begin{equation}
P_{\mathrm{rad}}(T)=~4\pi\displaystyle\int~\dfrac{\dfrac{2h{c}^{\textrm{{\tiny$2$}}}}{\lambda_{\textrm{{\tiny$e$}}}^{\textrm{{\tiny$5$}}}}~e_{\lambda_{\textrm{{\tiny$e$}}}}}{~\exp\left(\dfrac{hc}{\lambda_{\textrm{{\tiny$e$}}} k_{\mathrm{B}}T} \right)~-~1}~C_{\textrm{{\tiny{geo}}}}~\mathrm{d}\lambda_{\textrm{{\tiny$e$}}}
\label{eq:Prad}
\end{equation}
where \textit{h} is Planck's constant, \textit{c} is the speed of light in vacuum, and \textit{k}$_{\mathrm{B}}$ is Boltzmann's constant. A fundamental result in radiation transfer states that the thermal emissivity \textit{e}$_{\lambda}$ and the absorption efficiency \textit{Q}$_{\mathrm{abs}}$ of the NP are equal~\cite{Bohren2008}. Using this result, and inserting definitions~\eqref{eq:Pabs}~and~\eqref{eq:Prad} into~Eq.~\eqref{eq:energy_conservation}, we obtain the equation governing the NP temperature \textit{T} at the steady state:
\begin{equation}
Q_{\mathrm{abs}}^{\textrm{{\tiny$\beta$}}}\left(\lambda,\theta,T\right)I_{\mathrm{inc}}=\int\dfrac{C_{\textrm{{\tiny$1$}}}Q_{\mathrm{abs}}^{\textrm{{\tiny$\beta$}}}\left(\lambda_{\textrm{{\tiny$e$}}},\theta,T\right)}{\exp{\left(\dfrac{C_{\textrm{{\tiny$2$}}}}{\lambda_{\textrm{{\tiny$e$}}} T}\right)}-1}\dfrac{\mathrm{d}\lambda_{\textrm{{\tiny$e$}}}}{\lambda_{\textrm{{\tiny$e$}}}^{\textrm{{\tiny$5$}}}}~-~I_{\mathrm{rad}} \left(T_{\textrm{{\tiny{b}}}}\right)
\label{eq:Temperature}
\end{equation}
where we have used the notations~\textit{C}$_{\textrm{{\tiny$1$}}}$= $8\pi hc^{\textrm{{\tiny$2$}}}$, \textit{C}$_{\textrm{{\tiny$2$}}}$= $hc/k_{\mathrm{B}}$ for constants, and \textit{I}$_{\mathrm{rad}}\left(T_{\textrm{{\tiny$b$}}}\right)$ = $P_{\mathrm{rad}} \left(T_{\textrm{{\tiny{b}}}}\right)/C_{\textrm{{\tiny{geo}}}}$ 
refers to the power density emitted by the NP when just starting the illumination, {\it{i.e.}} at the bath temperature \textit{T}$_{\textrm{{\tiny{b}}}}$. We have supposed that the polychromatic light radiated by the NP ({\it{i.e.}} the thermal radiation) depends on the polarization $\beta$ and the angle of incidence $\theta$ of the external illumination.

For an incident illumination with fixed wavelength $\lambda$, polarization $\beta$, angle of incidence $\theta$, and power density \textit{I}$_{\mathrm{inc}}$, the steady-state temperature \textit{T} can be obtained numerically by solving Eq.~\eqref{eq:Temperature} using incremental-iterative procedure with adaptive stepsize \textit{dT}. The initial guess value used to start the iterative process has been chosen close to \textit{T}$_{\textrm{\tiny{b}}}$. We analytically calculated the NP absorption efficiency \textit{Q}$_{\mathrm{abs}}$ using an effective electric-dipole polarizability model (see Supplemental Material for full details~[\onlinecite{supp}]). This approach, in which the NP is treated in the electric-dipole approximation, requires a minimum gap distance \textit{z}$_{\textrm{\tiny{0}}}=2R$~[\onlinecite{Joulain2014}]. The scattering problem of a NP above a substrate is thus reduced to that of a NP with a dressed polarizability in a vacuum without a substrate. This fact justifies the use in our problem of the equality \textit{e}$_{\lambda}$=\textit{Q}$_{\mathrm{abs}}$ which is already established for a NP placed alone in a vacuum environment (see Sec. 4.7 of reference~\cite{Bohren2008}). In the numerical analyses presented below, we use a NP diameter $2R=50$ nm, and we assume that the NP does not undergo morphological changes~[\onlinecite{Setoura2014}] during the illumination. The bath temperature \textit{T}$_{\textrm{\tiny{b}}}$ is taken as a constant, unaffected by the heat dissipated from the NP into the surrounding medium, or by the external light absorbed by the substrate.

\section{Numerical results and discussion}
\label{sec:Results}

\subsection{Absorption efficiency of the NP}

The calculation of the temperature first requires an analysis of the absorption by the NP in front of a flat substrate. In Eq.~\eqref{eq:Temperature}, the only physical quantity that describes the electrodynamic response of the NP is the absorption efficiency \textit{Q}$_{\mathrm{abs}}$, that can be determined using an effective polarizability approach. In this approach, scattering and absorption of light by the NP in the presence of a substrate is described using a dressed polarizability, such that the induced dipole in the NP is proportional to the exciting field, the latter being defined as the field in the presence of the substrate but in absence of the NP. This way of modeling the problem leads to an analytic expression for \textit{Q}$_{\mathrm{abs}}$, which is appealing both for physical discussions and for a reduction of computer time. 

We consider here a gold NP above a gold substrate. The dielectric function of gold is described using a dispersion model that accounts for the temperature dependence, and for the finite size of the NP~[\onlinecite{Setoura2012}]. The details of the model are outlined in the Supplemental Material~[\onlinecite{supp}]. We set the temperature of the substrate at {\textit{T}}$_{\textrm{\tiny{b}}}$= 300 K, and choose an angle of incidence $\theta$=~45$^{\textrm{$\scriptscriptstyle{\circ}$}}$. The absorption efficiencies are calculated both for an incident electric field perpendicular to the plane of incidence (TE-polarized light), and for an incident electric field parallel to the plane of incidence (TM-polarized light).

\begin{figure}[!h]
\centering
\includegraphics[width=8cm]{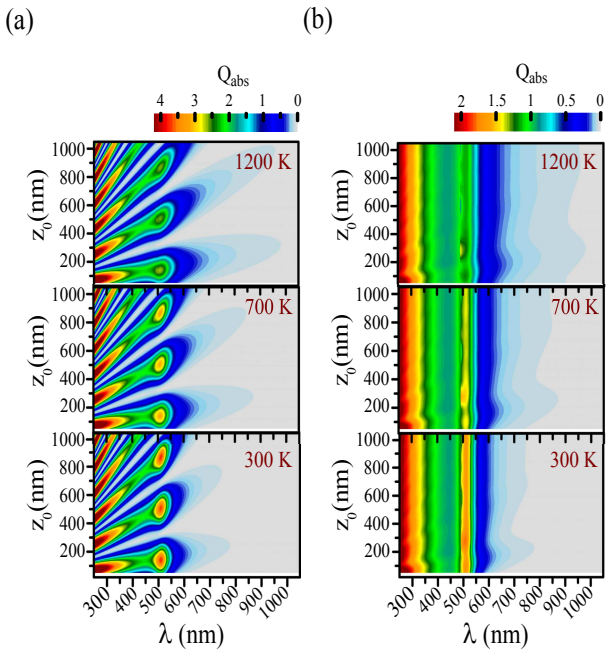}
\caption{Absorption efficiency of a single Au NP above a Au substrate for (a) TE- and (b) TM-polarization, versus the gap distance  \textit{z}$_{\textrm{\tiny{0}}}$ and the incident wavelength $\lambda$ for three NP temperatures {\textit{T}} = 300 K, 700 K, and 1200 K. NP diameter $2R=$ 50 nm, angle of incidence $\theta$=~45$^{\textrm{$\scriptscriptstyle{\circ}$}}$ and substrate temperature \textit{T}$_{\textrm{\tiny{b}}}$= 300 K.}
\label{fig2}
\end{figure}

Fig.~\ref{fig2} shows 2D maps of \textit{Q}$_{\textrm{abs}}$, plotted as a function of the wavelength of the incident light $\lambda$ and the NP-to-substrate gap distance \textit{z}$_{\tiny{0}}$ for three temperatures of the NP. The maps for TE-polarized incident light, presented in Fig.~\ref{fig2}\textcolor{blue}{(a)}, exhibit resonance branches including relatively high efficiencies. The maps in Fig.~\ref{fig2}\textcolor{blue}{(b)}, corresponding to TM-polarized incident light, are characterized by a quasi-uniform color for each wavelength all over the range of gap distances that has been studied. For both polarizations, the highest efficiencies are achieved in the UV range ($\lambda~\lesssim$ 350 nm) and around the surface plasmon resonance (SPR)~($\lambda\simeq$~506 nm). We point out that increasing the NP temperature has only one effect, which is the decrease of the efficiencies around the SPR. For instance, while the NP is heated up to 1200~K, the maximum efficiency at the SPR drops down to about 2.6  for TE polarization, corresponding to a decrease on the order of 30 \%. The decrease in efficiency due to the temperature is spectrally restricted to a narrow wavelength range around the SPR. This result is the same as that observed for a Au NP in free space. It should be noted that the absorption efficiency at the SPR in this case is significantly lower than the maximum value recorded in the presence of the Au substrate. In vacuum, \textit{Q}$_{\mathrm{abs}}$ at the SPR reaches a value of approximately 0.8 for a NP temperature of 1200 K (see Supplemental Material~Fig.~S2~[\onlinecite{supp}]). Therefore, the presence of the Au substrate generates a several fold increase in the absorption efficiency, producing a substantial antenna effect such that \textit{Q}$_{\mathrm{abs}}$ $>$ 1, or equivalently  \textit{C}$_{\mathrm{abs}}$ $>$ \textit{C}$_{\mathrm{geo}}$.

The origin of the different features observed in the maps in Fig.~\ref{fig2} can be analyzed in more detail. According to its definition, the absorption efficiency is not only a function of the effective polarizability $\alpha_{\mathrm{\tiny{eff}}}^{\mathrm{\tiny{E}}}$, but of the exciting field {\textbf{E}}$_{\mathrm{\tiny{exc}}}$ as well (see Supplemental Material~Eqs (S15)a,~(S15)b,~and (S5)~[\onlinecite{supp}]). The effective (or dressed) polarizability includes a modification to the original polarizability of the NP in free space. The correction term is proportional to the reflected part of the Green tensor ${\overset{\mathrm{\tiny{\bm\leftrightarrow}}}{\mathbf{G}}}^{\mathrm{\tiny{E}}}_{\mathrm{\tiny{refl}}}$ which can be expressed as a sum over propagating and evanescent plane waves (see Supplemental Material~Eqs~(S6),(S12)a, and (S12)b~[\onlinecite{supp}]). The evanescent waves are expected to contribute significantly in the regime $z_{\mathrm{\tiny{0}}}/\lambda \ll$ 1 (keeping in mind that the shortest distance $z_0=2R$). For the whole spectral range 250-1050 nm, the absorption efficiencies of the spherical Au NP above the Au substrate do not show any significant changes at short distances (slightly larger than 50 nm), regardless the polarization of the incident field. Changes are expected to appear for Au NP with smaller diameter (probing fields at shorter distances to the surface) or for other kind of NP with the same diameter, but with spectral resonances occuring at longer wavelengths. Actually, in the range of parameters considered here, the absorption of light by the NP above the substrate is weakly affected by the change in the polarizability (using the bare polarizability instead of the effective polarizability would hardly change the results). Since the bare polarizability is polarization-independent as the NP has a spherical shape, the most efficient absorption occurs at the same spectral ranges irrespective to the polarization, as seen in Fig.~\ref{fig2}. Over these spectral ranges, the Au NP alone already exhibits an efficient absorption (see Supplemental Material Fig.~S2~\cite{supp}). 

In addition to the effective polarizability, the absorption also depends on the exciting field. This field corresponds to the sum of the incident field and the field reflected by the flat surface (see Supplemental Material Eqs~(S13)a, (S16)a, (S16)b, (S17)a, and (S17)b \cite{supp}; see also Fig\ref{fig1}).
\begin{figure}[!h]
\centering
\includegraphics[width=7cm]{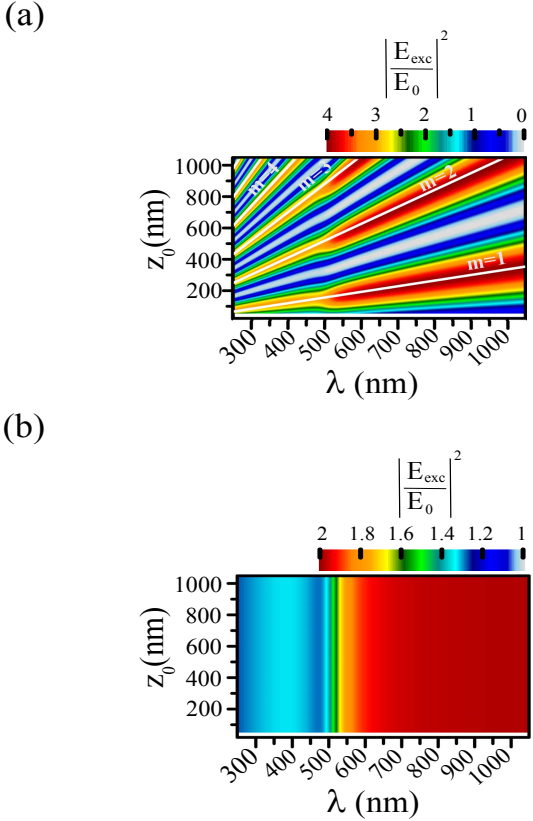}
\caption{Normalized electric-field intensity above the Au substrate in the absence of the NP (exciting field) for (a) TE- and (b) TM-polarization, versus the gap distance  \textit{z}$_{\textrm{\tiny{0}}}$ and the incident wavelength $\lambda$. White lines in (a) indicate the mean slope of the branches resulting from constructive interferences between the incident and reflected field (see text). Angle of incidence $\theta$=~45$^{\textrm{$\scriptscriptstyle{\circ}$}}$ and substrate temperature \textit{T}$_{\textrm{\tiny{b}}}$= 300 K.}
\label{fig3}
\end{figure}
We plot in Fig.~\ref{fig3} the intensity of the exciting field above the Au substrate in two dimensional color maps, considering the same spectral and gap distance ranges as in Fig.~\ref{fig2}. The map in Fig.~\ref{fig3}\textcolor{blue}{(a)} for TE polarization of the incident light shows branches associated with relatively intense exciting fields, while the map in Fig.~\ref{fig3}\textcolor{blue}{(b)} for TM polarization reveals a constant field intensity over the full range of distances for a given wavelength. These features are similar to those observed in the maps of \textit{Q}$_{\mathrm{abs}}$ in Fig.~\ref{fig2}. This similarity indicates that the absorption efficiency of the NP above the substrate is strongly modulated by the intensity of the exciting field, the modulation being observed only in the spectral range where the NP has a non-negligible absorption. The invariance of the field intensity with respect to the distance \textit{z}$_{\textrm{\tiny{0}}}$ for fixed wavelength, revealed in the map in Fig.~\ref{fig3}\textcolor{blue}{(b)}, is not a general behavior in TM polarization, but  is specific to the incidence angle $\theta$=~45$^{\textrm{$\scriptscriptstyle{\circ}$}}$. In the most general case, the field intensity map for TM polarization contains branches similar to that observed for TE polarization. The branches of the ($\lambda$,\textit{z}$_{\textrm{\tiny{0}}}$) map in Fig.~\ref{fig3}\textcolor{blue}{(a)} are aligned along straight thick lines. Thus, intense exciting fields are obtainable for particular ($\lambda$,\textit{z}$_{\textrm{\tiny{0}}}$) couples. The linear form of the branches results from the necessary condition for constructive interference between the incident and reflected fields. We refer to these branches as resonance branches. The mean slope of each resonance branche is shown as solid white line in the map in Fig.~\ref{fig3}\textcolor{blue}{(a)} using the following approximate formula $(2\pi/{\lambda})\cos\theta\left(2z_{\tiny{0}}\right)$~+~$\pi= m\, 2\pi$ where $m$ is a positive integer that refer to each branch. The intensity between two successive resonance branches can drop down to nearly zero, which indicates the occurrence of destructive interference. Note that the interference pattern explored here is specific to the Au substrate. The pattern may change substantially for another substrate material.
In summary, the results and the discussion above show that the NP absorption efficiency is strongly influenced by the exciting field, and can be tuned by changing either the NP-to-substrate gap distance or the incident wavelength.

\begin{figure}
\centering
\includegraphics[width=8cm]{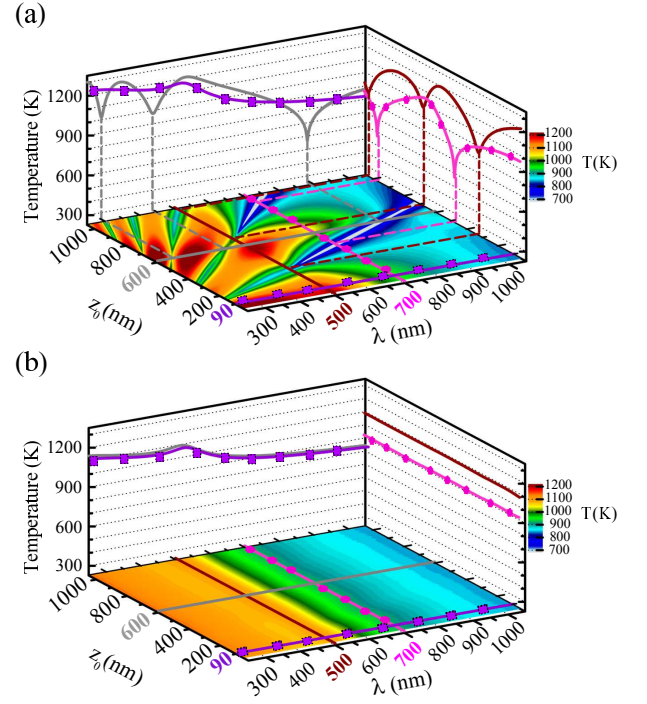}
\caption{Temperature of a single Au NP with diameter $2R=$~50 nm above a Au substrate versus the gap distance \textit{z}$_{\textrm{\tiny{0}}}$ and the excitation wavelength $\lambda$. Angle of incidence $\theta$=~45$^{\textrm{$\scriptscriptstyle{\circ}$}}$ and incident light power density \textit{I}$_{\mathrm{inc}}=1$ mW.cm$^{-\textrm{\tiny{2}}}$. The substrate temperature is \textit{T}$_{\textrm{\tiny{b}}}=300$ K. (a) TE polarized light, (b) TM polarized light. The plots also display the temperature versus $\lambda$ for two gap distances \textit{z}$_{\textrm{\tiny{0}}}=$~90 nm (squares connected by solid lines) and \textit{z}$_{\textrm{\tiny{0}}}=$~600 nm (solid lines), as well as the temperature versus \textit{z}$_{\textrm{\tiny{0}}}$ for two wavelengths $\lambda=$~500 nm (solid lines) and $\lambda=$~700 nm (circles connected by solid lines).} 
\label{fig4}
\end{figure}

\subsection{Steady-state temperature of the NP}
\label{subsec:TEMPCALC}

In this section, we present results of numerical calculations of the NP temperature based on Eq.~\eqref{eq:Temperature}. We compute the steady-state temperature for a Au NP with diameter $2R=50$ nm placed above a Au substrate, under CW illumination at a wavelength $\lambda$ with an intensity (power density) \textit{I}$_{\mathrm{inc}}= 1$ mW.cm$^{-\textrm{\tiny{2}}}$. As will be shown later, this low intensity value ensures that the NP temperature remains below the melting point. For the problem discussed here, we can find that the thermal resistance by radiation~\textit{Z}$_{\mathrm{rad}}$= $\Delta~T$/($P_{\mathrm{rad}}\left(T\right)~-~P_{\mathrm{rad}}\left(T_{\textrm{{\tiny{b}}}}\right)$)= $\Delta~T$/(\textit{Q}$_{\mathrm{abs}}$~\textit{I}$_{\mathrm{inc}}$~$\pi$\textit{R}$^{\mathrm{2}}$)~$\approx$~1.27~10$^{17}$~$\Delta~T$~(in Kelvin/Watt). It is worth to mention that regardless of the value of $\Delta~T$ ($\textgreater$ 1 K), the value of \textit{Z}$_{\mathrm{rad}}$ is always much smaller than the thermal resistance by conduction inside the NP \textit{Z}$_{\mathrm{cond}}$ $\approx$ 1/(2~$\kappa_{\textrm{\tiny{Au}}}~R$)= 6.3~10$^{4}$~Kelvin/Watt ($\kappa_{\textrm{\tiny{Au}}}$=317 W/(m.K) at $T=$ 300 K~\cite{Ho1972}). This is perfectly consistent with the assumption that the temperature inside the NP is uniform.

Figure~\ref{fig4} presents maps of the NP temperature versus the excitation wavelength $\lambda$ and the gap distance \textit{z}$_{\tiny{0}}$  for both incident polarizations. The 2D map in Fig.~\ref{fig4}\textcolor{blue}{(a)}, corresponding to TE polarization, exhibits branches with relatively high temperature values ranging from $T\sim$1000~K to $T\sim$1200~K, while for TM polarization, the 2D map in Fig.~\ref{fig4}\textcolor{blue}{(b)} shows that a given incident wavelength produces a NP temperature insensitive to the gap distance \textit{z}$_{\tiny{0}}$. These results follow naturally from the behavior of the absorption efficiency and of the exciting field intensity discussed in Figs.~\ref{fig2} and ~\ref{fig3}, respectively. Although the NP temperature is directly related to the intensity of the exciting field at a given wavelength $\lambda$ and at a given gap distance \textit{z}$_{\tiny{0}}$, the relation is not a simple proportionality factor. Indeed, an intense exciting field does not necessarily coincide with a high temperature value. For instance, the couples ($\lambda$,\textit{z}$_{\tiny{0}}$) in the map in Fig.~\ref{fig4}\textcolor{blue}{(a)} giving a high temperature (regions in dark-red color) do not all correspond to regions in the map in Fig.~\ref{fig2}\textcolor{blue}{(a)} for which the absorption efficiency is large. For example, when the NP is very close to the surface (\textit{z}$_{\tiny{0}}\simeq$ 50 nm), a relatively low value of \textit{Q}$_{\mathrm{abs}}$ can give rise to a relatively high temperature. Most often, however, when the absorption of the external illumination by the NP is so strong (i.e. high \textit{Q}$_{\textrm{abs}}$ value), the temperature achievable by the NP is so high, and {\it vice versa}. 

To go further in the analysis, 1D spectra are displayed beside the 2D maps in Fig.~\ref{fig4}. For instance, beside the map in Fig~\ref{fig4}\textcolor{blue}{(a)} the solid grey curve shows that the NP temperature oscillates with the incident wavelength $\lambda$ for a fixed gap distance \textit{z}$_{\textrm{\tiny{0}}}$=600 nm. One observes three temperature minima for different incident wavelengths. Furthermore, the solid dark-red curve beside the same map shows the temperature for a fixed wavelength $\lambda=500$~nm, indicating that the NP temperature exhibits maxima and minima when varying the gap distance \textit{z}$_{\textrm{\tiny{0}}}$. The same curves plotted beside the map in Fig.~\ref{fig4}\textcolor{blue}{(b)} are mainly featureless, and show relatively lower values of the temperature. Thus, with the angle of incidence chosen here, it is much easier to tune the temperature of the NP under TE illumination. Also note that for a given geometry and excitation wavelength, it is possible to cool or heat the NP by changing the polarization of the incident light. Again, these observations follow naturally from the behavior of the absorption efficiency of the NP and of the intensity of the exciting field above the substrate (see Figs~\ref{fig2} and~\ref{fig3}).

\subsubsection{Influence of the substrate}

It is interesting to discuss how the substrate influences the NP temperature. For the sake of comparison with the Au substrate studied so far, we consider a SiO$_{2}$ substrate. In the calculation, the dielectric function for bulk SiO$_{2}$ is taken from~Ref.~[\onlinecite{Palik1997}] and interpolated to cover all wavelengths of interest. All other parameters are kept identical to those used in the study with the Au substrate. 

\begin{figure}[!h]
\centering
\includegraphics[width=9cm]{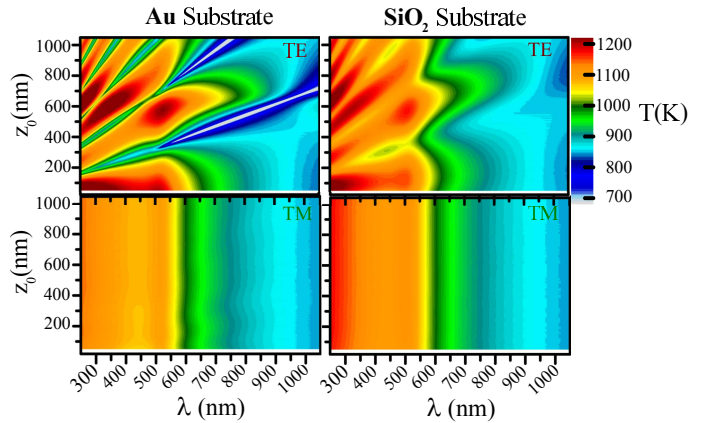}
\caption{Temperature of a single Au NP with diameter $2R=50$ nm above a Au substrate (left) and a SiO$_{2}$ substrate (right) versus the gap distance \textit{z}$_{\textrm{\tiny{0}}}$ and the excitation wavelength $\lambda$ for TE (top) and TM (bottom) incident polarizations. Angle of incidence $\theta$=
45$^{\textrm{$\scriptscriptstyle{\circ}$}}$ and incident light power density \textit{I}$_{\mathrm{inc}}=1$ mW.cm$^{-\textrm{\tiny{2}}}$. The substrate temperature is \textit{T}$_{\textrm{\tiny{b}}}=300$ K.}
\label{fig5}
\end{figure} 

\begin{figure}[!h]
\centering
\includegraphics[width=7cm]{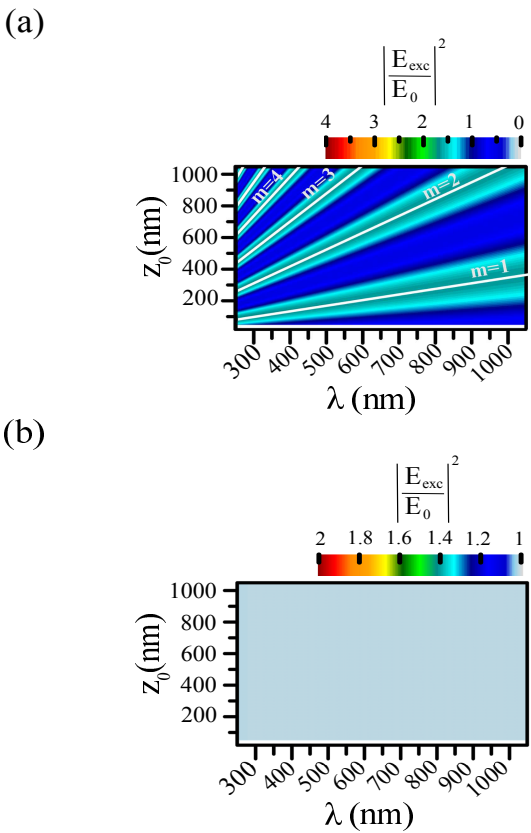}
\caption{same as Figure~\ref{fig3}, but for SiO$_{2}$ substrate.}
\label{fig6}
\end{figure}

In Fig.~\ref{fig5}, we plot the 2D maps of the NP temperature for the case of a SiO$_{2}$ substrate (right panel), for TE and TM polarizations of the incident field. For convenience, we have replotted the data for the Au substrate (left panel). The features in the maps for both cases are similar in terms of the appearance of branches specific to each polarization. For TM polarized light, the SiO$_{2}$ substrate produces much higher temperatures in the UV range ($\lambda$ $\lesssim$ 350 nm) compared to the Au substrate. For TE polarized light, the resonance branches in the maps are less distinguishable with the SiO$_{2}$ substrate. As we pointed out earlier, the steady-state temperature depends on the absorption efficiency of the external illumination by the NP \textit{Q}$_{\mathrm{abs}}$, which in turn is highly linked to the intensity of the exciting field above the substrate (i.e. the sum of incident and reflected fields in vacuum). For that reason, in order to understand the differences observed in the 2D maps we must examine the intensity of the exciting field above the SiO$_{2}$ substrate. We plot in Fig.~\ref{fig6} the corresponding 2D maps for both kinds of polarization. The map~\ref{fig6}\textcolor{blue}{(a)} for TE polarization shows branches comprising high intensity values ($\sim$ 1.6), which are however lower than those included in the branches of the map~\ref{fig3}\textcolor{blue}{(a)} of the gold substrate. Contrary to the case of gold substrate, we also remark that between two successive branches the exciting field intensity does not go down to zeros but to around 0.5. This reduced visibility of branches is that of the interference pattern above the substrate, which can be explained by the fact that over the studied wavelength range the SiO$_{2}$ substrate is considerably less reflective than the gold substrate. It clearly explains why the distinction between branches in the temperature map is less easy for SiO$_{2}$ substrate case. The map~\ref{fig6}\textcolor{blue}{(b)} for TM polarization reveals that the  intensity of the exciting field above the SiO$_{2}$ substrate is practically equal to the intensity of the incident field whatever the incident wavelength and the NP-to-substrate gap distance. This result implies that the absorption efficiency spectrum corresponds to that of the NP in vacuum without substrate (see Supplemental Material Fig. S2). Thus, the absorption is much stronger in UV range, which allows the achievement of highest temperature values in that wavelength range. 

\subsubsection{Influence of the NP material}

One can also expect the temperature to be strongly dependent on the NP material, in particular due to changes in the absorption resonance spectrum. It is interesting to study NPs exhbiting surface phonon polariton resonance (SPhPR) in the mid-infrared (MIR) spectral region~[\onlinecite{Caldwell2015}], such as a NP made of SiC in the presence of a SiC or SiO$_{2}$ substrate. To perform the calculations, we have used bulk dielectric functions for both the NP and the substrate. The data are taken (and interpolated) from the temperature-dependent data calculated by Meneses {\it et al.} from spectral reflectance measurements~[\onlinecite{Meneses2014},\onlinecite{Joulain2015}] (see Supplemental Material Figs S5(a) and S6(a)~\cite{supp}). All other parameters are the same as in the previous calculations with the Au NP.

\begin{figure}[!h]
\centering
\includegraphics[width=9cm]{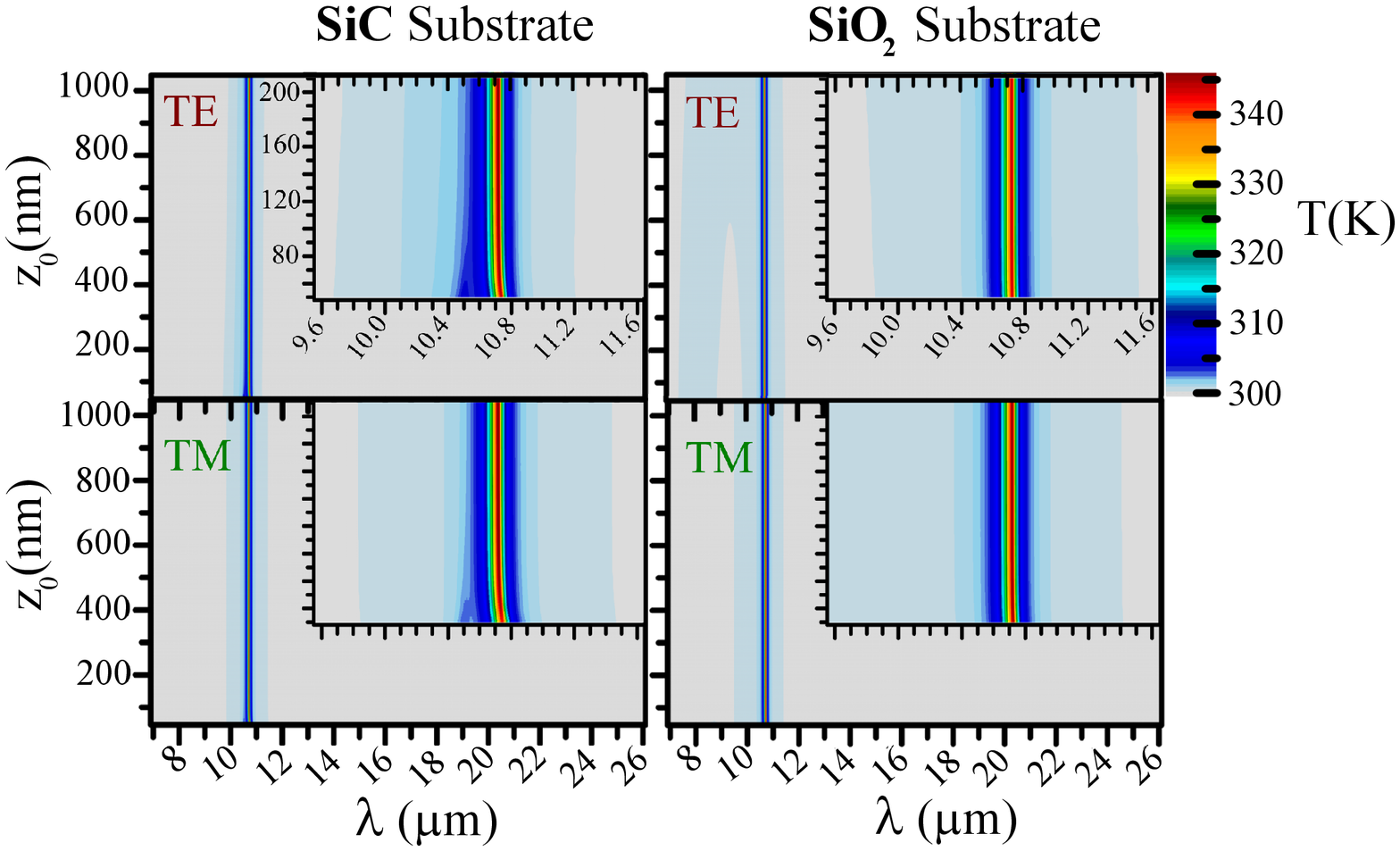}
\caption{Temperature of a single SiC NP with diameter $2R = 50$ nm above a SiC substrate (left) and a SiO$_{2}$ substrate (right) versus the gap distance \textit{z}$_{\textrm{\tiny{0}}}$ and the excitation wavelength $\lambda$ for TE (top) and TM (bottom) incident polarizations. Angle of incidence $\theta$=~45$^{\textrm{$\scriptscriptstyle{\circ}$}}$ and incident light power density \textit{I}$_{\mathrm{inc}}=1$ mW.cm$^{-\textrm{\tiny{2}}}$. The substrate temperature is \textit{T}$_{\textrm{\tiny{b}}}=300$ K. The insets show a zoom of the map around the surface phonon polariton resonance.}
\label{fig7}
\end{figure}

We show in Fig.~\ref{fig7} maps of the temperature of a SiC NP versus the gap distance \textit{z}$_{\textrm{\tiny{0}}}$ and the excitation wavelength $\lambda$ above a SiC substrate (left) and a SiO$_{2}$ substrate (right), and for TE (top) and TM (bottom) incident polarizations.  All maps display an increase in temperature in a narrow spectral range, corresponding to the SPhPR ($\sim$10.7 $\mu$m). Outside the resonance, the temperature remains close to the bath temperature ($\simeq 300$ K). This behavior is very different from that observed for the Au NP (see Figs~\ref{fig4} and~\ref{fig5}), where a significant increase in temperature occurs over a broader range of wavelengths. For the SiC NP, the absorption spectrum is characterized by a sharp peak associated with the SPhPR (see Supplemental Material Fig.~S5(b)), which explains the thermal behavior. Moreover, the maximum temperature increase is~$\sim$45~K, which is significantly lower than the value~$\sim$900~K obtained with the Au NP (see Fig.~\ref{fig4}~or~Fig.~\ref{fig5}). Although not shown for the sake of brevity, for a SiC NP (with temperature T= 300 K) placed above a SiC substrate, the absorption efficiency under TE polarized illumination reaches a maximum value~$\sim$2 at large gap distances. This value is comparable to the maximum efficiency value of $\sim$2.6 achieved at SPR with a Au NP of temperature T= 1200~K placed above a Au substrate (see Fig.~\ref{fig2}). The difference in temperature increase between SiC and Au NP is thus due to the spectral range at which the thermal emission occurs. The spectral range of the thermal emission process is driven by Plank's function [see Eq.~\eqref{eq:Prad}], with larger values in the MIR than in the visible. As a consequence, for a SiC NP, a little increase in temperature allows to completely dissipate the absorbed power~(see Eq.~\ref{eq:energy_conservation}).

Let us now examine the absorption behavior at smaller gap distances. The inset of each map in Fig.~\ref{fig7} depicts a zoom around the SPhPR for distances \textit{z}$_{\textrm{\tiny{0}}}<210$ nm. We see that for the SiO$_{2}$ substrate (right panel) the absorption does not undergo any change even at short distance. However, for the SiC substrate (left panel), slight changes are observed at distances smaller than 100 nm. In fact, the spectral position of the NP temperature peak (shown in dark-red on the 2D map) shifts smoothly towards longer wavelength with the moving of the NP closer to the substrate surface. This shift is a result of the interaction between two spectrally close optical resonances (surface phonon polaritons resonances), the optical resonance of the SiC NP and that of the SiC substrate. This near-field optical interaction affects the NP temperature through the modification of the absorption efficiency of the external illumination by the NP. The probing of such interaction here at NP-to-substrate gap distance larger than 50 nm (in contrast to the case of gold NP above a gold substrate) can be directly related to the fact that the absorption spectrum (as well as the thermal emission spectrum) of the SiC NP is located at the mid-infrared, which renders the near-field regime well-satisfied ($z_{\mathrm{\tiny{0}}}/\lambda \ll$ 1).
Finally, let us note that the SiC NP temperature is weakly dependent on the polarization of the incident light in the range of parameters considered here.

\subsection{Critical melting intensity for a Au NP}

The steady-state temperature of the NP depends directly on the incident intensity (power density) \textit{I}$_{\mathrm{inc}}$ (see Eq.~\eqref{eq:Temperature}). The approach presented in section~\ref{sec:Method} applies only to temperatures below the NP melting point. The enthalpy of fusion should be taken into account in Eq.~\eqref{eq:Temperature} for higher temperatures~[\onlinecite{Setoura2014},\onlinecite{Bisker2012}]. Moreover, the dielectric function data used in the numerical simulations are valid for the solid phase only. In particular, the dispersion data used for the dielectric function of Au describe exclusively the temperature dependence for solid Au~[\onlinecite{Setoura2012}]~(see Supplemental Material for details about this model~\cite{supp}). To determine the range of validity of the model, it is necessary to evaluate the critical intensity that brings the NP temperature up to the melting point. This calculation is interesting as well from the point of view of thermal management, since it will show that the substrate and the incident wavelength substantially change the critical intensity.

We have chosen to consider the case of a Au NP with diameter $2R=50$ nm above a Au substrate, and an illumination at normal incidence. With this illumination, the TE and TM polarizations are equivalent, and the absorption efficiency {\textit{Q}}$_{\textrm{abs}}$ is independent on polarization. We set the NP temperature \textit{T} equal to the melting temperature of the Au NP, using \textit{T}$=1337$~K that corresponds to bulk Au~[\onlinecite{Guenther2014}]. The temperature of the substrate is {\textit{T}}$_{\textrm{\tiny{b}}}=300$~K. 

\begin{figure}
\centering
\includegraphics[width=8cm]{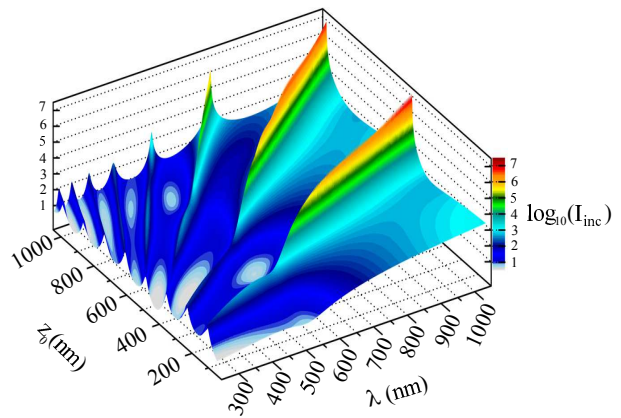}
\caption{Decimal logarithm of the critical incident intensity {\textit{I}}$_{\mathrm{inc}}$ needed to heat a Au NP above a Au substrate up to the melting point (melting temperature 1337 K) versus the gap distance \textit{z}$_{\textrm{\tiny{0}}}$ and the excitation wavelength $\lambda$. {\textit{I}}$_{\mathrm{inc}}$ is in unit mW.cm$^{-\textrm{\tiny{2}}}$. The NP diameter is $2R=50$ nm, the substrate temperature is \textit{T}$_{\textrm{\tiny{b}}}=300$~K, and the incident wave is at normal incidence ($\theta$=~0$^{\textrm{$\scriptscriptstyle{\circ}$}}$).} 
\label{fig8}
\end{figure}

In Fig.~\ref{fig8}, we plot the decimal logarithm of the critical intensity (in unit mW.cm$^{-2}$) versus the incident wavelength $\lambda$ and the gap distance {\textit{z}}$_{\tiny{0}}$. The intensity values range from a few mW.cm$^{-2}$ to about 10$^{4}$ W.cm$^{-2}$. This shows that for a Au NP and a Au substrate, keeping the incident intensity to a value {\textit{I}}$_{\mathrm{inc}}=1$~mW.cm$^{-2}$, as has been done throughout this study, prevents the NP temperature to reach the melting point for all the explored values of ($\lambda$,\textit{z}$_{\tiny{0}}$). The highest observed critical intensity ($\sim$10$^{4}$ W.cm$^{-2}$) is lower than the critical intensity calculated for a Au NP of the same size in air in free space, the values of {\textit{I}}$_{\textrm{\tiny{inc}}}$ in this varying approximately between~3.10$^{4}$ and 3.10$^{7}$ W.cm$^{-2}$ (see Supplemental Material Fig.~S7~\cite{supp}). Setoura \textit{et al.}~[\onlinecite{Setoura2013}] found that a CW laser at 488 nm with an intensity of 2.10$^{5}$ W.cm$^{-2}$ increases the temperature of a Au NP of 100 nm diameter in air supported on glass substrate up to $\sim$ 600~K. The NP temperature is estimated from dark-field light scattering measurements. It should be noted that a value of {\textit{I}}$_{\tiny{inc}}\sim$5.10$^{5}$ W.cm$^{-2}$ is theoretically sufficient to heat a 50 nm diameter Au NP in air up to its melting point. 

\section{Conclusion}
\label{sec:concl}

In summary, we have developed a model to calculate the steady-state temperature of a single NP above a flat substrate under CW external illumination, and with thermal radiation as the only heat loss channel. The interaction between the NP and the substrate is described using an effective electric-dipole polarizability. We showed that the absorption of the Au NP, which exhibits SPR in the visible range, depends on the exciting intensity pattern generated above the substrate by the interference between the incident and reflected waves. Interestingly, the Au NP temperature exhibits interference modulations while varying the NP-to-substrate gap distance or the wavelength of the incident light. The temperature increase is tunable by either changing the incident-light polarization, or the substrate, two parameters that control the interference modulations. In contrast, for SiC NP which supports SPhPR in the mid-infrared, we find that the temperature is weakly dependent on the gap distance, the polarization, or the substrate, in particular for distances smaller than 1 $\mu$m.  To increase the temperature of such NP, the wavelength of the incident light must be selected to match the SPhPR. This model provides a tool for the thermal characterization of micro or nanoscale architectures involving flat substrates and NPs or optical antennas. Further development could include an additional conductive heat loss channel between the NP and the substrate, and work in this direction is in progress.
\begin{acknowledgments}

We acknowledge the financial support from the French National Research Agency (ANR) through NATO project grant (ANR-13-BS10-0013). H.K. is grateful to University of Poitiers for post-doctoral fellowship. The calculations were performed using high performance computing facilities at PPrime Institute and M\'{e}socentre SPIN Calcul,  University of Poitiers. 
\end{acknowledgments}
\bibliography{references}

\begin{thebibliography}{44}%
\makeatletter
\providecommand \@ifxundefined [1]{%
 \@ifx{#1\undefined}
}%
\providecommand \@ifnum [1]{%
 \ifnum #1\expandafter \@firstoftwo
 \else \expandafter \@secondoftwo
 \fi
}%
\providecommand \@ifx [1]{%
 \ifx #1\expandafter \@firstoftwo
 \else \expandafter \@secondoftwo
 \fi
}%
\providecommand \natexlab [1]{#1}%
\providecommand \enquote  [1]{``#1''}%
\providecommand \bibnamefont  [1]{#1}%
\providecommand \bibfnamefont [1]{#1}%
\providecommand \citenamefont [1]{#1}%
\providecommand \href@noop [0]{\@secondoftwo}%
\providecommand \href [0]{\begingroup \@sanitize@url \@href}%
\providecommand \@href[1]{\@@startlink{#1}\@@href}%
\providecommand \@@href[1]{\endgroup#1\@@endlink}%
\providecommand \@sanitize@url [0]{\catcode `\\12\catcode `\$12\catcode
  `\&12\catcode `\#12\catcode `\^12\catcode `\_12\catcode `\%12\relax}%
\providecommand \@@startlink[1]{}%
\providecommand \@@endlink[0]{}%
\providecommand \url  [0]{\begingroup\@sanitize@url \@url }%
\providecommand \@url [1]{\endgroup\@href {#1}{\urlprefix }}%
\providecommand \urlprefix  [0]{URL }%
\providecommand \Eprint [0]{\href }%
\providecommand \doibase [0]{http://dx.doi.org/}%
\providecommand \selectlanguage [0]{\@gobble}%
\providecommand \bibinfo  [0]{\@secondoftwo}%
\providecommand \bibfield  [0]{\@secondoftwo}%
\providecommand \translation [1]{[#1]}%
\providecommand \BibitemOpen [0]{}%
\providecommand \bibitemStop [0]{}%
\providecommand \bibitemNoStop [0]{.\EOS\space}%
\providecommand \EOS [0]{\spacefactor3000\relax}%
\providecommand \BibitemShut  [1]{\csname bibitem#1\endcsname}%
\let\auto@bib@innerbib\@empty
\bibitem [{\citenamefont {Govorov}\ \emph {et~al.}(2006)\citenamefont
  {Govorov}, \citenamefont {Zhang}, \citenamefont {Skeini}, \citenamefont
  {Richardson}, \citenamefont {Lee},\ and\ \citenamefont
  {Kotov}}]{Govorov2006}%
  \BibitemOpen
  \bibfield  {author} {\bibinfo {author} {\bibfnamefont {A.~O.}\ \bibnamefont
  {Govorov}}, \bibinfo {author} {\bibfnamefont {W.}~\bibnamefont {Zhang}},
  \bibinfo {author} {\bibfnamefont {T.}~\bibnamefont {Skeini}}, \bibinfo
  {author} {\bibfnamefont {H.}~\bibnamefont {Richardson}}, \bibinfo {author}
  {\bibfnamefont {J.}~\bibnamefont {Lee}}, \ and\ \bibinfo {author}
  {\bibfnamefont {N.~A.}\ \bibnamefont {Kotov}},\ }\href {\doibase
  10.1007/s11671-006-9015-7} {\bibfield  {journal} {\bibinfo  {journal}
  {Nanoscale Res. Lett.}\ }\textbf {\bibinfo {volume} {1(1)}},\ \bibinfo
  {pages} {84} (\bibinfo {year} {2006})}\BibitemShut {NoStop}%
\bibitem [{\citenamefont {Baffou}\ and\ \citenamefont
  {Rigneault}(2011)}]{Baffou2011}%
  \BibitemOpen
  \bibfield  {author} {\bibinfo {author} {\bibfnamefont {G.}~\bibnamefont
  {Baffou}}\ and\ \bibinfo {author} {\bibfnamefont {H.}~\bibnamefont
  {Rigneault}},\ }\href {\doibase 10.1103/PhysRevB.84.035415} {\bibfield
  {journal} {\bibinfo  {journal} {Phys. Rev. B}\ }\textbf {\bibinfo {volume}
  {84}},\ \bibinfo {pages} {035415} (\bibinfo {year} {2011})}\BibitemShut
  {NoStop}%
\bibitem [{\citenamefont {Carlson}\ \emph {et~al.}(2011)\citenamefont
  {Carlson}, \citenamefont {Khan},\ and\ \citenamefont
  {Richardson}}]{Carlson2011}%
  \BibitemOpen
  \bibfield  {author} {\bibinfo {author} {\bibfnamefont {M.~T.}\ \bibnamefont
  {Carlson}}, \bibinfo {author} {\bibfnamefont {A.}~\bibnamefont {Khan}}, \
  and\ \bibinfo {author} {\bibfnamefont {H.~H.}\ \bibnamefont {Richardson}},\
  }\href {\doibase 10.1021/nl103938u} {\bibfield  {journal} {\bibinfo
  {journal} {Nano Lett.}\ }\textbf {\bibinfo {volume} {11}},\ \bibinfo {pages}
  {1061} (\bibinfo {year} {2011})}\BibitemShut {NoStop}%
\bibitem [{\citenamefont {Setoura}\ \emph {et~al.}(2012)\citenamefont
  {Setoura}, \citenamefont {Werner},\ and\ \citenamefont
  {Hashimoto}}]{Setoura2012}%
  \BibitemOpen
  \bibfield  {author} {\bibinfo {author} {\bibfnamefont {K.}~\bibnamefont
  {Setoura}}, \bibinfo {author} {\bibfnamefont {D.}~\bibnamefont {Werner}}, \
  and\ \bibinfo {author} {\bibfnamefont {S.}~\bibnamefont {Hashimoto}},\ }\href
  {\doibase 10.1021/jp304271d} {\bibfield  {journal} {\bibinfo  {journal} {J.
  Phys. Chem. C}\ }\textbf {\bibinfo {volume} {116}},\ \bibinfo {pages} {15458}
  (\bibinfo {year} {2012})}\BibitemShut {NoStop}%
\bibitem [{\citenamefont {Boisselier}\ and\ \citenamefont
  {Astruc}(2009)}]{Boisselier2009}%
  \BibitemOpen
  \bibfield  {author} {\bibinfo {author} {\bibfnamefont {E.}~\bibnamefont
  {Boisselier}}\ and\ \bibinfo {author} {\bibfnamefont {D.}~\bibnamefont
  {Astruc}},\ }\href {\doibase 10.1039/B806051G} {\bibfield  {journal}
  {\bibinfo  {journal} {Chem. Soc. Rev.}\ }\textbf {\bibinfo {volume} {38}},\
  \bibinfo {pages} {1759} (\bibinfo {year} {2009})}\BibitemShut {NoStop}%
\bibitem [{\citenamefont {Jaque}\ \emph {et~al.}(2014)\citenamefont {Jaque},
  \citenamefont {Martinez~Maestro}, \citenamefont {del Rosal}, \citenamefont
  {Haro-Gonzalez}, \citenamefont {Benayas}, \citenamefont {Plaza},
  \citenamefont {Martin~Rodriguez},\ and\ \citenamefont
  {Garcia~Sole}}]{Jaque2014}%
  \BibitemOpen
  \bibfield  {author} {\bibinfo {author} {\bibfnamefont {D.}~\bibnamefont
  {Jaque}}, \bibinfo {author} {\bibfnamefont {L.}~\bibnamefont
  {Martinez~Maestro}}, \bibinfo {author} {\bibfnamefont {B.}~\bibnamefont {del
  Rosal}}, \bibinfo {author} {\bibfnamefont {P.}~\bibnamefont {Haro-Gonzalez}},
  \bibinfo {author} {\bibfnamefont {A.}~\bibnamefont {Benayas}}, \bibinfo
  {author} {\bibfnamefont {J.~L.}\ \bibnamefont {Plaza}}, \bibinfo {author}
  {\bibfnamefont {E.}~\bibnamefont {Martin~Rodriguez}}, \ and\ \bibinfo
  {author} {\bibfnamefont {J.}~\bibnamefont {Garcia~Sole}},\ }\href {\doibase
  10.1039/C4NR00708E} {\bibfield  {journal} {\bibinfo  {journal} {Nanoscale}\
  }\textbf {\bibinfo {volume} {6}},\ \bibinfo {pages} {9494} (\bibinfo {year}
  {2014})}\BibitemShut {NoStop}%
\bibitem [{\citenamefont {Cao}\ \emph {et~al.}(2007)\citenamefont {Cao},
  \citenamefont {Barsic}, \citenamefont {Guichard},\ and\ \citenamefont
  {Brongersma}}]{Cao2007}%
  \BibitemOpen
  \bibfield  {author} {\bibinfo {author} {\bibfnamefont {L.}~\bibnamefont
  {Cao}}, \bibinfo {author} {\bibfnamefont {D.~N.}\ \bibnamefont {Barsic}},
  \bibinfo {author} {\bibfnamefont {A.~R.}\ \bibnamefont {Guichard}}, \ and\
  \bibinfo {author} {\bibfnamefont {M.~L.}\ \bibnamefont {Brongersma}},\ }\href
  {\doibase 10.1021/nl0722370} {\bibfield  {journal} {\bibinfo  {journal} {Nano
  Lett.}\ }\textbf {\bibinfo {volume} {7}},\ \bibinfo {pages} {3523} (\bibinfo
  {year} {2007})}\BibitemShut {NoStop}%
\bibitem [{\citenamefont {Schuller}\ \emph {et~al.}(2009)\citenamefont
  {Schuller}, \citenamefont {Taubner},\ and\ \citenamefont
  {Brongersma}}]{Schuller2009}%
  \BibitemOpen
  \bibfield  {author} {\bibinfo {author} {\bibfnamefont {J.~A.}\ \bibnamefont
  {Schuller}}, \bibinfo {author} {\bibfnamefont {T.}~\bibnamefont {Taubner}}, \
  and\ \bibinfo {author} {\bibfnamefont {M.~L.}\ \bibnamefont {Brongersma}},\
  }\href {\doibase 10.1038/nphoton.2009.188} {\bibfield  {journal} {\bibinfo
  {journal} {Nat. Photon.}\ }\textbf {\bibinfo {volume} {3}},\ \bibinfo {pages}
  {658} (\bibinfo {year} {2009})}\BibitemShut {NoStop}%
\bibitem [{\citenamefont {Centini}\ \emph {et~al.}(2015)\citenamefont
  {Centini}, \citenamefont {Benedetti}, \citenamefont {Larciprete},
  \citenamefont {Belardini}, \citenamefont {Li~Voti}, \citenamefont
  {Bertolotti},\ and\ \citenamefont {Sibilia}}]{Centini2015}%
  \BibitemOpen
  \bibfield  {author} {\bibinfo {author} {\bibfnamefont {M.}~\bibnamefont
  {Centini}}, \bibinfo {author} {\bibfnamefont {A.}~\bibnamefont {Benedetti}},
  \bibinfo {author} {\bibfnamefont {M.~C.}\ \bibnamefont {Larciprete}},
  \bibinfo {author} {\bibfnamefont {A.}~\bibnamefont {Belardini}}, \bibinfo
  {author} {\bibfnamefont {R.}~\bibnamefont {Li~Voti}}, \bibinfo {author}
  {\bibfnamefont {M.}~\bibnamefont {Bertolotti}}, \ and\ \bibinfo {author}
  {\bibfnamefont {C.}~\bibnamefont {Sibilia}},\ }\href {\doibase
  10.1103/PhysRevB.92.205411} {\bibfield  {journal} {\bibinfo  {journal} {Phys.
  Rev. B}\ }\textbf {\bibinfo {volume} {92}},\ \bibinfo {pages} {205411}
  (\bibinfo {year} {2015})}\BibitemShut {NoStop}%
\bibitem [{\citenamefont {Mulet}\ \emph {et~al.}(2001)\citenamefont {Mulet},
  \citenamefont {Joulain}, \citenamefont {Carminati},\ and\ \citenamefont
  {Greffet}}]{Mulet2001}%
  \BibitemOpen
  \bibfield  {author} {\bibinfo {author} {\bibfnamefont {J.-P.}\ \bibnamefont
  {Mulet}}, \bibinfo {author} {\bibfnamefont {K.}~\bibnamefont {Joulain}},
  \bibinfo {author} {\bibfnamefont {R.}~\bibnamefont {Carminati}}, \ and\
  \bibinfo {author} {\bibfnamefont {J.-J.}\ \bibnamefont {Greffet}},\ }\href
  {\doibase 10.1063/1.1370118} {\bibfield  {journal} {\bibinfo  {journal}
  {Appl. Phys. Lett.}\ }\textbf {\bibinfo {volume} {78}},\ \bibinfo {pages}
  {2931} (\bibinfo {year} {2001})}\BibitemShut {NoStop}%
\bibitem [{\citenamefont {Narayanaswamy}\ \emph {et~al.}(2008)\citenamefont
  {Narayanaswamy}, \citenamefont {Shen},\ and\ \citenamefont
  {Chen}}]{Narayanaswamy2008}%
  \BibitemOpen
  \bibfield  {author} {\bibinfo {author} {\bibfnamefont {A.}~\bibnamefont
  {Narayanaswamy}}, \bibinfo {author} {\bibfnamefont {S.}~\bibnamefont {Shen}},
  \ and\ \bibinfo {author} {\bibfnamefont {G.}~\bibnamefont {Chen}},\ }\href
  {\doibase 10.1103/PhysRevB.78.115303} {\bibfield  {journal} {\bibinfo
  {journal} {Phys. Rev. B}\ }\textbf {\bibinfo {volume} {78}},\ \bibinfo
  {pages} {115303} (\bibinfo {year} {2008})}\BibitemShut {NoStop}%
\bibitem [{\citenamefont {Chapuis}\ \emph {et~al.}(2008)\citenamefont
  {Chapuis}, \citenamefont {Laroche}, \citenamefont {Volz},\ and\ \citenamefont
  {Greffet}}]{Chapuis20081}%
  \BibitemOpen
  \bibfield  {author} {\bibinfo {author} {\bibfnamefont {P.-O.}\ \bibnamefont
  {Chapuis}}, \bibinfo {author} {\bibfnamefont {M.}~\bibnamefont {Laroche}},
  \bibinfo {author} {\bibfnamefont {S.}~\bibnamefont {Volz}}, \ and\ \bibinfo
  {author} {\bibfnamefont {J.-J.}\ \bibnamefont {Greffet}},\ }\href {\doibase
  10.1103/PhysRevB.77.125402} {\bibfield  {journal} {\bibinfo  {journal} {Phys.
  Rev. B}\ }\textbf {\bibinfo {volume} {77}},\ \bibinfo {pages} {125402}
  (\bibinfo {year} {2008})}\BibitemShut {NoStop}%
\bibitem [{\citenamefont {Biehs}\ and\ \citenamefont
  {Greffet}(2010)}]{Biehs2010}%
  \BibitemOpen
  \bibfield  {author} {\bibinfo {author} {\bibfnamefont {S.-A.}\ \bibnamefont
  {Biehs}}\ and\ \bibinfo {author} {\bibfnamefont {J.-J.}\ \bibnamefont
  {Greffet}},\ }\href {\doibase 10.1103/PhysRevB.81.245414} {\bibfield
  {journal} {\bibinfo  {journal} {Phys. Rev. B}\ }\textbf {\bibinfo {volume}
  {81}},\ \bibinfo {pages} {245414} (\bibinfo {year} {2010})}\BibitemShut
  {NoStop}%
\bibitem [{\citenamefont {Tschikin}\ \emph {et~al.}(2012)\citenamefont
  {Tschikin}, \citenamefont {Biehs}, \citenamefont {Rosa},\ and\ \citenamefont
  {Ben-Abdallah}}]{Tschikin2012}%
  \BibitemOpen
  \bibfield  {author} {\bibinfo {author} {\bibfnamefont {M.}~\bibnamefont
  {Tschikin}}, \bibinfo {author} {\bibfnamefont {S.-A.}\ \bibnamefont {Biehs}},
  \bibinfo {author} {\bibfnamefont {F.}~\bibnamefont {Rosa}}, \ and\ \bibinfo
  {author} {\bibfnamefont {P.}~\bibnamefont {Ben-Abdallah}},\ }\href {\doibase
  10.1140/epjb/e2012-30219-7} {\bibfield  {journal} {\bibinfo  {journal} {Eur.
  Phys. J. B}\ }\textbf {\bibinfo {volume} {85}},\ \bibinfo {pages} {1}
  (\bibinfo {year} {2012})}\BibitemShut {NoStop}%
\bibitem [{\citenamefont {Miller}\ \emph {et~al.}(2015)\citenamefont {Miller},
  \citenamefont {Johnson},\ and\ \citenamefont {Rodriguez}}]{Miller2015}%
  \BibitemOpen
  \bibfield  {author} {\bibinfo {author} {\bibfnamefont {O.~D.}\ \bibnamefont
  {Miller}}, \bibinfo {author} {\bibfnamefont {S.~G.}\ \bibnamefont {Johnson}},
  \ and\ \bibinfo {author} {\bibfnamefont {A.~W.}\ \bibnamefont {Rodriguez}},\
  }\href {\doibase 10.1103/PhysRevLett.115.204302} {\bibfield  {journal}
  {\bibinfo  {journal} {Phys. Rev. Lett.}\ }\textbf {\bibinfo {volume} {115}},\
  \bibinfo {pages} {204302} (\bibinfo {year} {2015})}\BibitemShut {NoStop}%
\bibitem [{\citenamefont {Joulain}\ \emph {et~al.}(2014)\citenamefont
  {Joulain}, \citenamefont {Ben-Abdallah}, \citenamefont {Chapuis},
  \citenamefont {Wilde}, \citenamefont {Babuty},\ and\ \citenamefont
  {Henkel}}]{Joulain2014}%
  \BibitemOpen
  \bibfield  {author} {\bibinfo {author} {\bibfnamefont {K.}~\bibnamefont
  {Joulain}}, \bibinfo {author} {\bibfnamefont {P.}~\bibnamefont
  {Ben-Abdallah}}, \bibinfo {author} {\bibfnamefont {P.-O.}\ \bibnamefont
  {Chapuis}}, \bibinfo {author} {\bibfnamefont {Y.~D.}\ \bibnamefont {Wilde}},
  \bibinfo {author} {\bibfnamefont {A.}~\bibnamefont {Babuty}}, \ and\ \bibinfo
  {author} {\bibfnamefont {C.}~\bibnamefont {Henkel}},\ }\href {\doibase
  10.1016/j.jqsrt.2013.12.006} {\bibfield  {journal} {\bibinfo  {journal} {J.
  Quant. Spectrosc. Radiat. Transfer}\ }\textbf {\bibinfo {volume} {136}},\
  \bibinfo {pages} {1 } (\bibinfo {year} {2014})}\BibitemShut {NoStop}%
\bibitem [{\citenamefont {Jones}\ and\ \citenamefont
  {Raschke}(2012)}]{Jones2012}%
  \BibitemOpen
  \bibfield  {author} {\bibinfo {author} {\bibfnamefont {A.~C.}\ \bibnamefont
  {Jones}}\ and\ \bibinfo {author} {\bibfnamefont {M.~B.}\ \bibnamefont
  {Raschke}},\ }\href {\doibase 10.1021/nl204201g} {\bibfield  {journal}
  {\bibinfo  {journal} {Nano Lett.}\ }\textbf {\bibinfo {volume} {12}},\
  \bibinfo {pages} {1475} (\bibinfo {year} {2012})}\BibitemShut {NoStop}%
\bibitem [{\citenamefont {Babuty}\ \emph {et~al.}(2013)\citenamefont {Babuty},
  \citenamefont {Joulain}, \citenamefont {Chapuis}, \citenamefont {Greffet},\
  and\ \citenamefont {De~Wilde}}]{Babuty2013}%
  \BibitemOpen
  \bibfield  {author} {\bibinfo {author} {\bibfnamefont {A.}~\bibnamefont
  {Babuty}}, \bibinfo {author} {\bibfnamefont {K.}~\bibnamefont {Joulain}},
  \bibinfo {author} {\bibfnamefont {P.-O.}\ \bibnamefont {Chapuis}}, \bibinfo
  {author} {\bibfnamefont {J.-J.}\ \bibnamefont {Greffet}}, \ and\ \bibinfo
  {author} {\bibfnamefont {Y.}~\bibnamefont {De~Wilde}},\ }\href {\doibase
  10.1103/PhysRevLett.110.146103} {\bibfield  {journal} {\bibinfo  {journal}
  {Phys. Rev. Lett.}\ }\textbf {\bibinfo {volume} {110}},\ \bibinfo {pages}
  {146103} (\bibinfo {year} {2013})}\BibitemShut {NoStop}%
\bibitem [{\citenamefont {Loke}\ \emph {et~al.}(2011)\citenamefont {Loke},
  \citenamefont {Menguc},\ and\ \citenamefont {Nieminen}}]{Loke2011}%
  \BibitemOpen
  \bibfield  {author} {\bibinfo {author} {\bibfnamefont {V.~L.}\ \bibnamefont
  {Loke}}, \bibinfo {author} {\bibfnamefont {M.~P.}\ \bibnamefont {Menguc}}, \
  and\ \bibinfo {author} {\bibfnamefont {T.~A.}\ \bibnamefont {Nieminen}},\
  }\href {\doibase 10.1016/j.jqsrt.2011.03.012} {\bibfield  {journal} {\bibinfo
   {journal} {J. Quant. Spectrosc. Radiat. Transfer}\ }\textbf {\bibinfo
  {volume} {112}},\ \bibinfo {pages} {1711 } (\bibinfo {year}
  {2011})}\BibitemShut {NoStop}%
\bibitem [{\citenamefont {Yurkin}\ and\ \citenamefont
  {Huntemann}(2015)}]{Yurkin2015}%
  \BibitemOpen
  \bibfield  {author} {\bibinfo {author} {\bibfnamefont {M.~A.}\ \bibnamefont
  {Yurkin}}\ and\ \bibinfo {author} {\bibfnamefont {M.}~\bibnamefont
  {Huntemann}},\ }\href {\doibase 10.1021/acs.jpcc.5b09271} {\bibfield
  {journal} {\bibinfo  {journal} {J. Phys. Chem. C}\ }\textbf {\bibinfo
  {volume} {119}},\ \bibinfo {pages} {29088} (\bibinfo {year}
  {2015})}\BibitemShut {NoStop}%
\bibitem [{\citenamefont {Edalatpour}\ and\ \citenamefont
  {Francoeur}(2016)}]{Edalatpour2016}%
  \BibitemOpen
  \bibfield  {author} {\bibinfo {author} {\bibfnamefont {S.}~\bibnamefont
  {Edalatpour}}\ and\ \bibinfo {author} {\bibfnamefont {M.}~\bibnamefont
  {Francoeur}},\ }\href {\doibase 10.1103/PhysRevB.94.045406} {\bibfield
  {journal} {\bibinfo  {journal} {Phys. Rev. B}\ }\textbf {\bibinfo {volume}
  {94}},\ \bibinfo {pages} {045406} (\bibinfo {year} {2016})}\BibitemShut
  {NoStop}%
\bibitem [{\citenamefont {Waxenegger}\ \emph {et~al.}(2015)\citenamefont
  {Waxenegger}, \citenamefont {Tr\"{u}gler},\ and\ \citenamefont
  {Hohenester}}]{Waxenegger2015}%
  \BibitemOpen
  \bibfield  {author} {\bibinfo {author} {\bibfnamefont {J.}~\bibnamefont
  {Waxenegger}}, \bibinfo {author} {\bibfnamefont {A.}~\bibnamefont
  {Tr\"{u}gler}}, \ and\ \bibinfo {author} {\bibfnamefont {U.}~\bibnamefont
  {Hohenester}},\ }\href {\doibase 10.1016/j.cpc.2015.03.023} {\bibfield
  {journal} {\bibinfo  {journal} {Comp. Phys. Commun.}\ }\textbf {\bibinfo
  {volume} {193}},\ \bibinfo {pages} {138 } (\bibinfo {year}
  {2015})}\BibitemShut {NoStop}%
\bibitem [{\citenamefont {Videen}(1991)}]{Videen1991}%
  \BibitemOpen
  \bibfield  {author} {\bibinfo {author} {\bibfnamefont {G.}~\bibnamefont
  {Videen}},\ }\href {\doibase 10.1364/JOSAA.8.000483} {\bibfield  {journal}
  {\bibinfo  {journal} {J. Opt. Soc. Am. A}\ }\textbf {\bibinfo {volume} {8}},\
  \bibinfo {pages} {483} (\bibinfo {year} {1991})}\BibitemShut {NoStop}%
\bibitem [{\citenamefont {Videen}(1992)}]{Videen1992}%
  \BibitemOpen
  \bibfield  {author} {\bibinfo {author} {\bibfnamefont {G.}~\bibnamefont
  {Videen}},\ }\href {\doibase 10.1364/JOSAA.9.000844} {\bibfield  {journal}
  {\bibinfo  {journal} {J. Opt. Soc. Am. A}\ }\textbf {\bibinfo {volume} {9}},\
  \bibinfo {pages} {844} (\bibinfo {year} {1992})}\BibitemShut {NoStop}%
\bibitem [{\citenamefont {Videen}(1993)}]{Videen1993}%
  \BibitemOpen
  \bibfield  {author} {\bibinfo {author} {\bibfnamefont {G.}~\bibnamefont
  {Videen}},\ }\href {\doibase 10.1364/JOSAA.10.000110} {\bibfield  {journal}
  {\bibinfo  {journal} {J. Opt. Soc. Am. A}\ }\textbf {\bibinfo {volume}
  {10}},\ \bibinfo {pages} {110} (\bibinfo {year} {1993})}\BibitemShut
  {NoStop}%
\bibitem [{\citenamefont {Videen}(1995)}]{Videen1995}%
  \BibitemOpen
  \bibfield  {author} {\bibinfo {author} {\bibfnamefont {G.}~\bibnamefont
  {Videen}},\ }\href {\doibase 10.1016/0030-4018(94)00668-K} {\bibfield
  {journal} {\bibinfo  {journal} {Opt. Commun.}\ }\textbf {\bibinfo {volume}
  {115}},\ \bibinfo {pages} {1 } (\bibinfo {year} {1995})}\BibitemShut
  {NoStop}%
\bibitem [{\citenamefont {Videen}(2000)}]{Videen2000}%
  \BibitemOpen
  \bibfield  {author} {\bibinfo {author} {\bibfnamefont {G.}~\bibnamefont
  {Videen}},\ }\enquote {\bibinfo {title} {Light scattering from a sphere near
  a plane interface},}\ in\ \href {\doibase 10.1007/3-540-46614-2_5} {\emph
  {\bibinfo {booktitle} {Light Scattering from Microstructures. Lecture Notes
  in Physics}}},\ Vol.\ \bibinfo {volume} {534},\ \bibinfo {editor} {edited by\
  \bibinfo {editor} {\bibfnamefont {F.}~\bibnamefont {Moreno}}\ and\ \bibinfo
  {editor} {\bibfnamefont {F.}~\bibnamefont {Gonz{\'a}lez}}}\ (\bibinfo
  {publisher} {Springer Berlin Heidelberg},\ \bibinfo {address} {Berlin,
  Heidelberg},\ \bibinfo {year} {2000})\ Chap.~\bibinfo {chapter} {5}, pp.\
  \bibinfo {pages} {81--96}\BibitemShut {NoStop}%
\bibitem [{\citenamefont {Fucile}\ \emph {et~al.}(1997)\citenamefont {Fucile},
  \citenamefont {Denti}, \citenamefont {Borghese}, \citenamefont {Saija},\ and\
  \citenamefont {Sindoni}}]{Fucile1997}%
  \BibitemOpen
  \bibfield  {author} {\bibinfo {author} {\bibfnamefont {E.}~\bibnamefont
  {Fucile}}, \bibinfo {author} {\bibfnamefont {P.}~\bibnamefont {Denti}},
  \bibinfo {author} {\bibfnamefont {F.}~\bibnamefont {Borghese}}, \bibinfo
  {author} {\bibfnamefont {R.}~\bibnamefont {Saija}}, \ and\ \bibinfo {author}
  {\bibfnamefont {O.~I.}\ \bibnamefont {Sindoni}},\ }\href {\doibase
  10.1364/JOSAA.14.001505} {\bibfield  {journal} {\bibinfo  {journal} {J. Opt.
  Soc. Am. A}\ }\textbf {\bibinfo {volume} {14}},\ \bibinfo {pages} {1505}
  (\bibinfo {year} {1997})}\BibitemShut {NoStop}%
\bibitem [{\citenamefont {Wriedt}\ and\ \citenamefont
  {Doicu}(1998)}]{Wriedt1998}%
  \BibitemOpen
  \bibfield  {author} {\bibinfo {author} {\bibfnamefont {T.}~\bibnamefont
  {Wriedt}}\ and\ \bibinfo {author} {\bibfnamefont {A.}~\bibnamefont {Doicu}},\
  }\href {\doibase 10.1016/S0030-4018(98)00099-6} {\bibfield  {journal}
  {\bibinfo  {journal} {Opt. Commun.}\ }\textbf {\bibinfo {volume} {152}},\
  \bibinfo {pages} {376 } (\bibinfo {year} {1998})}\BibitemShut {NoStop}%
\bibitem [{\citenamefont {Mackowski}(2008)}]{Mackowski2008}%
  \BibitemOpen
  \bibfield  {author} {\bibinfo {author} {\bibfnamefont {D.~W.}\ \bibnamefont
  {Mackowski}},\ }\href {\doibase 10.1016/j.jqsrt.2007.08.024} {\bibfield
  {journal} {\bibinfo  {journal} {J. Quant. Spectrosc. Radiat. Transfer}\
  }\textbf {\bibinfo {volume} {109}},\ \bibinfo {pages} {770 } (\bibinfo {year}
  {2008})}\BibitemShut {NoStop}%
\bibitem [{\citenamefont {Lerm\'{e}}\ \emph {et~al.}(2013)\citenamefont
  {Lerm\'{e}}, \citenamefont {Bonnet}, \citenamefont {Broyer}, \citenamefont
  {Cottancin}, \citenamefont {Manchon},\ and\ \citenamefont
  {Pellarin}}]{Lerme2013}%
  \BibitemOpen
  \bibfield  {author} {\bibinfo {author} {\bibfnamefont {J.}~\bibnamefont
  {Lerm\'{e}}}, \bibinfo {author} {\bibfnamefont {C.}~\bibnamefont {Bonnet}},
  \bibinfo {author} {\bibfnamefont {M.}~\bibnamefont {Broyer}}, \bibinfo
  {author} {\bibfnamefont {E.}~\bibnamefont {Cottancin}}, \bibinfo {author}
  {\bibfnamefont {D.}~\bibnamefont {Manchon}}, \ and\ \bibinfo {author}
  {\bibfnamefont {M.}~\bibnamefont {Pellarin}},\ }\href {\doibase
  10.1021/jp3121963} {\bibfield  {journal} {\bibinfo  {journal} {J. Phys. Chem.
  C}\ }\textbf {\bibinfo {volume} {117}},\ \bibinfo {pages} {6383} (\bibinfo
  {year} {2013})}\BibitemShut {NoStop}%
\bibitem [{\citenamefont {Hillenbrand}\ \emph {et~al.}(2002)\citenamefont
  {Hillenbrand}, \citenamefont {Taubner},\ and\ \citenamefont
  {Keilmann}}]{Hillenbrand2002}%
  \BibitemOpen
  \bibfield  {author} {\bibinfo {author} {\bibfnamefont {R.}~\bibnamefont
  {Hillenbrand}}, \bibinfo {author} {\bibfnamefont {T.}~\bibnamefont
  {Taubner}}, \ and\ \bibinfo {author} {\bibfnamefont {F.}~\bibnamefont
  {Keilmann}},\ }\href {\doibase 10.1038/nature00899} {\bibfield  {journal}
  {\bibinfo  {journal} {Nature}\ }\textbf {\bibinfo {volume} {418}},\ \bibinfo
  {pages} {159} (\bibinfo {year} {2002})}\BibitemShut {NoStop}%
\bibitem [{\citenamefont {Miroshnichenko}\ \emph {et~al.}(2015)\citenamefont
  {Miroshnichenko}, \citenamefont {Evlyukhin}, \citenamefont {Kivshar},\ and\
  \citenamefont {Chichkov}}]{Miroshnichenko2015}%
  \BibitemOpen
  \bibfield  {author} {\bibinfo {author} {\bibfnamefont {A.~E.}\ \bibnamefont
  {Miroshnichenko}}, \bibinfo {author} {\bibfnamefont {A.~B.}\ \bibnamefont
  {Evlyukhin}}, \bibinfo {author} {\bibfnamefont {Y.~S.}\ \bibnamefont
  {Kivshar}}, \ and\ \bibinfo {author} {\bibfnamefont {B.~N.}\ \bibnamefont
  {Chichkov}},\ }\href {\doibase 10.1021/acsphotonics.5b00117} {\bibfield
  {journal} {\bibinfo  {journal} {ACS Photon.}\ }\textbf {\bibinfo {volume}
  {2}},\ \bibinfo {pages} {1423} (\bibinfo {year} {2015})}\BibitemShut
  {NoStop}%
\bibitem [{\citenamefont {Bohren}\ and\ \citenamefont
  {Huffman}(2008)}]{Bohren2008}%
  \BibitemOpen
  \bibfield  {author} {\bibinfo {author} {\bibfnamefont {C.~F.}\ \bibnamefont
  {Bohren}}\ and\ \bibinfo {author} {\bibfnamefont {D.~R.}\ \bibnamefont
  {Huffman}},\ }\href {\doibase 10.1002/9783527618156} {\emph {\bibinfo {title}
  {Absorption and Scattering of Light by Small Particles}}}\ (\bibinfo
  {publisher} {John Wiley \& Sons},\ \bibinfo {address} {New York},\ \bibinfo
  {year} {2008})\BibitemShut {NoStop}%
\bibitem [{sup()}]{supp}%
  \BibitemOpen
  \href@noop {} {}\bibinfo {note} {See Supplemental Material at
  \url{http://link.aps.org/supplemental/10.1103/PhysRevB.XX.XXXXXX} for
  effective dipole-polarizability approach, dispersion model of Au NP
  dielectric function, and calculation results for NP absorption in abscence of
  substrate, and light intensity for melting Au NP placed in air.}\BibitemShut
  {Stop}%
\bibitem [{\citenamefont {Setoura}\ \emph {et~al.}(2014)\citenamefont
  {Setoura}, \citenamefont {Okada},\ and\ \citenamefont
  {Hashimoto}}]{Setoura2014}%
  \BibitemOpen
  \bibfield  {author} {\bibinfo {author} {\bibfnamefont {K.}~\bibnamefont
  {Setoura}}, \bibinfo {author} {\bibfnamefont {Y.}~\bibnamefont {Okada}}, \
  and\ \bibinfo {author} {\bibfnamefont {S.}~\bibnamefont {Hashimoto}},\ }\href
  {\doibase 10.1039/C4CP03733B} {\bibfield  {journal} {\bibinfo  {journal}
  {Phys. Chem. Chem. Phys.}\ }\textbf {\bibinfo {volume} {16}},\ \bibinfo
  {pages} {26938} (\bibinfo {year} {2014})}\BibitemShut {NoStop}%
\bibitem [{\citenamefont {Ho}\ \emph {et~al.}(1972)\citenamefont {Ho},
  \citenamefont {Powell},\ and\ \citenamefont {Liley}}]{Ho1972}%
  \BibitemOpen
  \bibfield  {author} {\bibinfo {author} {\bibfnamefont {C.~Y.}\ \bibnamefont
  {Ho}}, \bibinfo {author} {\bibfnamefont {R.~W.}\ \bibnamefont {Powell}}, \
  and\ \bibinfo {author} {\bibfnamefont {P.~E.}\ \bibnamefont {Liley}},\ }\href
  {\doibase http://dx.doi.org/10.1063/1.3253100} {\bibfield  {journal}
  {\bibinfo  {journal} {J. Phys. Chem. Ref. Data}\ }\textbf {\bibinfo {volume}
  {1}},\ \bibinfo {pages} {279} (\bibinfo {year} {1972})}\BibitemShut {NoStop}%
\bibitem [{\citenamefont {Philipp}(1997)}]{Palik1997}%
  \BibitemOpen
  \bibfield  {author} {\bibinfo {author} {\bibfnamefont {H.~R.}\ \bibnamefont
  {Philipp}},\ }in\ \href {\doibase 10.1016/B978-012544415-6.50038-8} {\emph
  {\bibinfo {booktitle} {Handbook of Optical Constants of Solids}}},\
  Vol.~\bibinfo {volume} {I},\ \bibinfo {editor} {edited by\ \bibinfo {editor}
  {\bibfnamefont {E.~D.}\ \bibnamefont {Palik}}}\ (\bibinfo  {publisher}
  {Academic Press},\ \bibinfo {year} {1997})\ pp.\ \bibinfo {pages}
  {749--763}\BibitemShut {NoStop}%
\bibitem [{\citenamefont {Caldwell}\ \emph {et~al.}(2015)\citenamefont
  {Caldwell}, \citenamefont {Lucas}, \citenamefont {Vincenzo}, \citenamefont
  {Igor}, \citenamefont {L.}, \citenamefont {A.},\ and\ \citenamefont
  {J.}}]{Caldwell2015}%
  \BibitemOpen
  \bibfield  {author} {\bibinfo {author} {\bibfnamefont {J.~D.}\ \bibnamefont
  {Caldwell}}, \bibinfo {author} {\bibfnamefont {L.}~\bibnamefont {Lucas}},
  \bibinfo {author} {\bibfnamefont {G.}~\bibnamefont {Vincenzo}}, \bibinfo
  {author} {\bibfnamefont {V.}~\bibnamefont {Igor}}, \bibinfo {author}
  {\bibfnamefont {R.~T.}\ \bibnamefont {L.}}, \bibinfo {author} {\bibfnamefont
  {M.~S.}\ \bibnamefont {A.}}, \ and\ \bibinfo {author} {\bibfnamefont {G.~O.}\
  \bibnamefont {J.}},\ }\href
  {www.degruyter.com/view/j/nanoph.2015.4.issue-1/nanoph-2014-0003/nanoph-2014-0003.xml}
  {\bibfield  {journal} {\bibinfo  {journal} {Nanophotonics}\ }\textbf
  {\bibinfo {volume} {4}},\  (\bibinfo {year} {2015})}\BibitemShut {NoStop}%
\bibitem [{\citenamefont {Meneses}\ \emph {et~al.}(2014)\citenamefont
  {Meneses}, \citenamefont {Eckes}, \citenamefont {del Campo},\ and\
  \citenamefont {Echegut}}]{Meneses2014}%
  \BibitemOpen
  \bibfield  {author} {\bibinfo {author} {\bibfnamefont {D.~D.~S.}\
  \bibnamefont {Meneses}}, \bibinfo {author} {\bibfnamefont {M.}~\bibnamefont
  {Eckes}}, \bibinfo {author} {\bibfnamefont {L.}~\bibnamefont {del Campo}}, \
  and\ \bibinfo {author} {\bibfnamefont {P.}~\bibnamefont {Echegut}},\ }\href
  {http://stacks.iop.org/0953-8984/26/i=25/a=255402} {\bibfield  {journal}
  {\bibinfo  {journal} {J. Phys. Condens. Matter}\ }\textbf {\bibinfo {volume}
  {26}},\ \bibinfo {pages} {255402} (\bibinfo {year} {2014})}\BibitemShut
  {NoStop}%
\bibitem [{\citenamefont {Joulain}\ \emph {et~al.}(2015)\citenamefont
  {Joulain}, \citenamefont {Ezzahri}, \citenamefont {Drevillon}, \citenamefont
  {Rousseau},\ and\ \citenamefont {Meneses}}]{Joulain2015}%
  \BibitemOpen
  \bibfield  {author} {\bibinfo {author} {\bibfnamefont {K.}~\bibnamefont
  {Joulain}}, \bibinfo {author} {\bibfnamefont {Y.}~\bibnamefont {Ezzahri}},
  \bibinfo {author} {\bibfnamefont {J.}~\bibnamefont {Drevillon}}, \bibinfo
  {author} {\bibfnamefont {B.}~\bibnamefont {Rousseau}}, \ and\ \bibinfo
  {author} {\bibfnamefont {D.~D.~S.}\ \bibnamefont {Meneses}},\ }\href
  {\doibase 10.1364/OE.23.0A1388} {\bibfield  {journal} {\bibinfo  {journal}
  {Opt. Express}\ }\textbf {\bibinfo {volume} {23}},\ \bibinfo {pages} {A1388}
  (\bibinfo {year} {2015})}\BibitemShut {NoStop}%
\bibitem [{\citenamefont {Bisker}\ and\ \citenamefont
  {Yelin}(2012)}]{Bisker2012}%
  \BibitemOpen
  \bibfield  {author} {\bibinfo {author} {\bibfnamefont {G.}~\bibnamefont
  {Bisker}}\ and\ \bibinfo {author} {\bibfnamefont {D.}~\bibnamefont {Yelin}},\
  }\href {\doibase 10.1364/JOSAB.29.001383} {\bibfield  {journal} {\bibinfo
  {journal} {J. Opt. Soc. Am. B}\ }\textbf {\bibinfo {volume} {29}},\ \bibinfo
  {pages} {1383} (\bibinfo {year} {2012})}\BibitemShut {NoStop}%
\bibitem [{\citenamefont {Guenther}\ and\ \citenamefont
  {Guillon}(2014)}]{Guenther2014}%
  \BibitemOpen
  \bibfield  {author} {\bibinfo {author} {\bibfnamefont {G.}~\bibnamefont
  {Guenther}}\ and\ \bibinfo {author} {\bibfnamefont {O.}~\bibnamefont
  {Guillon}},\ }\href {\doibase 10.1007/s10853-014-8544-1} {\bibfield
  {journal} {\bibinfo  {journal} {J. Mater. Sci.}\ }\textbf {\bibinfo {volume}
  {49}},\ \bibinfo {pages} {7915} (\bibinfo {year} {2014})}\BibitemShut
  {NoStop}%
\bibitem [{\citenamefont {Setoura}\ \emph {et~al.}(2013)\citenamefont
  {Setoura}, \citenamefont {Okada}, \citenamefont {Werner},\ and\ \citenamefont
  {Hashimoto}}]{Setoura2013}%
  \BibitemOpen
  \bibfield  {author} {\bibinfo {author} {\bibfnamefont {K.}~\bibnamefont
  {Setoura}}, \bibinfo {author} {\bibfnamefont {Y.}~\bibnamefont {Okada}},
  \bibinfo {author} {\bibfnamefont {D.}~\bibnamefont {Werner}}, \ and\ \bibinfo
  {author} {\bibfnamefont {S.}~\bibnamefont {Hashimoto}},\ }\href {\doibase
  10.1021/nn402863s} {\bibfield  {journal} {\bibinfo  {journal} {ACS Nano}\
  }\textbf {\bibinfo {volume} {7}},\ \bibinfo {pages} {7874} (\bibinfo {year}
  {2013})}\BibitemShut {NoStop}%
\end{thebibliography}%

\end{document}